\documentstyle[12pt,epsfig]{article}
\begin{document} 
\vspace*{-1in} 
\renewcommand{\thefootnote}{\fnsymbol{footnote}} 
\begin{flushright} 
TIFR/TH/99-20\\
May 1999\\ 
hep-ph/9905395\\
\end{flushright} 
\vskip 65pt 
\begin{center} 
{\Large \bf Probing Large Extra Dimensions Using Top Production
in Photon-Photon Collisions}\\
\vspace{8mm} 
{\bf 
Prakash Mathews\footnote{prakash@theory.tifr.res.in}, 
P. Poulose\footnote{poulose@theory.tifr.res.in}, 
K.~Sridhar\footnote{sridhar@theory.tifr.res.in}
}\\ 
\vspace{10pt} 
{\sf Department of Theoretical Physics, Tata Institute of 
Fundamental Research,\\  
Homi Bhabha Road, Bombay 400 005, India. 
} 
 
\vspace{80pt} 
{\bf ABSTRACT} 
\end{center} 
\vskip12pt 
\noindent Theories with large extra dimensions predict an infinite
tower of Kaluza-Klein states in the 1~TeV range, which can consequently
have significant implications for experimental observables. One
such observable, which gets affected by the exchange of spin-2
Kaluza-Klein particles, is the $t \bar t$ production cross-section
in photon-photon collisions at NLC energies. We study this process
and obtain bounds on the effective quantum gravity scale $M_S$ 
between 1600 and 5400 GeV (depending on the centre-of-mass energy).
We show that the use of polarisation will further strengthen these
bounds.

\setcounter{footnote}{0} 
\renewcommand{\thefootnote}{\arabic{footnote}} 
 
\vfill 
\clearpage 
\setcounter{page}{1} 
\pagestyle{plain}
\noindent 
Recently there has been tremendous interest in the study of the 
effects of large compact extra dimensions \cite{dimo}. In such
a scenario, the effects of gravity could become large at very low scales 
($\sim$~TeV). Starting from a string theory in 10 dimensions
\cite{dimo2, shiu}, the effective low-energy theory is obtained by
compactifying to (3+1) dimensions, in such a way that $n$ of these extra 
dimensions are compactified to a common scale $R$ which is large,
while the remaining dimensions are compactified to scales of the order 
of the inverse Planck scale. In such a scenario, the Standard Model (SM) 
particles (corresponding to open strings) live on a 3-brane and are, 
therefore, confined to the $(3+1)$-dimensional spacetime. On the other hand, 
the gravitons (corresponding to closed strings) propagate in the 
$(4+n)$-dimensional bulk. The low-energy scale $M_S$ is related to
the Planck scale by \cite{dimo}
\begin{equation} 
M^2_{\rm P}=M_{S}^{n+2} R^n ~,
\label{e1} 
\end{equation} 
so that we can choose $M_S$ to be of the order of a TeV and thus get around 
the hierarchy problem. It then follows that $R=10^{32/n -19}$~m, and so we 
find that $M_S$ can be arranged to be a TeV for any value $n > 1$. Deviations
from classical gravity can become apparent at these surprisingly low values 
of energy. For example, for $n=2$ the compactified dimensions
are of the order of 1 mm, just below the experimentally tested region
for the validity of classical gravity and within the possible
reach of ongoing experiments \cite{gravexp}.
In fact, it has been shown \cite{dimo4} that it is 
possible to construct a phenomenologically viable scenario with large
extra dimensions, which can survive the existing astrophysical and 
cosmological constraints. There have also been several interesting
studies of the implications of the large Kaluza-Klein dimensions for 
gauge coupling unification \cite{dienes}.
For some early papers on large Kaluza-Klein dimensions, see Ref.~\cite{anto, 
taylor} and for recent investigations on different aspects of the
TeV scale quantum gravity scenario and related ideas, see Ref.~\cite{related}.

Below the scale $M_S$ \cite{sundrum, grw, hlz}, we have an
effective theory with an infinite tower of massive Kaluza-Klein
states. which contain spin-2, spin-1 and spin-0 excitations. 
The only states that contribute to low-energy phenomenology in
an important manner are the spin-2 Kaluza-Klein states i.e. the
infinite tower of massive graviton states in the effective theory. 
For graviton momenta smaller than the scale $M_S$, the gravitons 
couple to the SM fields via a (four-dimensional) induced metric $g_{\mu \nu}$. 
The interactions of the SM particles with the graviton, $G_{\mu\nu}$, 
can be derived from the following Lagrangian \cite{hlz, grw} :
\begin{equation} 
{\cal L}=-{1 \over \bar M_P} G_{\mu \nu}^{(j)}T^{\mu\nu} ~,
\label{e2} 
\end{equation} 
where $j$ labels the Kaluza-Klein mode, $\bar M_P=M_P/\sqrt{8\pi}$
and $T^{\mu\nu}$ is the energy-momentum tensor. In spite of being
suppressed by $1/\bar M_P$, the effects of these couplings are 
significant for collider energies because of the fact that the
sum over the tower of graviton states cancels the dependence on $\bar M_P$ 
and, instead, provides a suppression of the order of $M_S$. 

There have been several studies exploring the consequences of the above
effective Lagrangian for experimental observables at high-energy 
colliders. The gravitons can be directly produced at 
$e^+ e^-$ or hadron colliders leading to spectacular single photon + 
missing energy or monojet + missing energy signatures \cite{mpp, keung}. 
Non-observation of these modes yields bounds which are 
around 500 GeV to 1.2 TeV at LEP2 \cite{mpp, keung} and around 600 GeV 
to 750 GeV at Tevatron (for $n$ between 2 and 6) \cite{mpp}. 
These studies have been extended to
the Large Hadron Collider (LHC) and to high-energy $e^+ e^-$ collisions
at the Next Linear Collider (NLC). Other than looking for the production
of gravitons, one can also study the effects of the exchange of
virtual gravitons in the intermediate state on experimental observables.
Graviton exchange in $e^+ e^- \rightarrow f \bar f$ and in high-mass 
dilepton production \cite{hewett,gmr}, in $t \bar t$ production \cite{us} 
at the Tevatron and the LHC, in deep-inelastic scattering at HERA 
\cite{us2}, and in jet production at the Tevatron \cite{us3} have 
been studied. Virtual effects in dilepton production at 
Tevatron yields a bound of around 950 GeV \cite{hewett} while a
more thorough statistical analysis increases this bound by about 100~GeV
\cite{gmr}, $t \bar t$ 
production at Tevatron yields a bound of about 650 GeV \cite{us}, while 
from deep-inelastic scattering a bound of 550 GeV results \cite{us2}. 
Jet production at the Tevatron yields strong bounds of about 1.2 TeV 
\cite{us3}.
At LHC, it is expected that $t \bar t$ production can be used to explore 
a range of $M_S$ values upto 4~TeV \cite{us}. More recently, fermion
pair production and gauge boson production in $e^+ e^-$ collisions at LEP2 
and NLC and in $\gamma \gamma$ collisions at the NLC \cite{rizzo, 
agashe, soni, davoudiasl, lee} have been studied. Associated production of 
gravitons with gauge bosons and virtual effects in gauge boson pair production
at hadron colliders have also been studied \cite{balazs}.
Diphoton signals and
global lepton-quark neutral current constraints have also
been studied \cite{cheung}. There have also been papers 
discussing the implications of the large dimensions for higgs
production \cite{rizzo2, xhe} and electroweak precision observables 
\cite{precision}. Astrophysical constraints, like bounds from energy loss 
for supernovae cores, have also been discussed \cite{astro}.

The present work concentrates on studying the effects of large extra
dimensions in top production in photon-photon collisions at the
NLC. The photons are produced in the Compton back-scattering of a 
highly monochromatic low-energy laser beam off a high energy electron beam
\cite{nlc}. Control over the $e^-$ and
laser beam parameters allow for control over the parameters
of the photon-photon subprocess. The physics potential of the
NLC in the photon-photon mode is manifold and these experiments
are planned over several steps of energy spanning the range between
500 GeV and 1.5 TeV. These experiments also provide a great degree
of precision because of the relatively clean initial state, and
indeed the degree of precision can be enhanced by using polarised 
initial electrons and laser beams. 
The precision that is possible in these experiments
and the high energies that are planned to be accessed provide
an ideal testing ground of the SM and a very effective
probe of possible physics that may lie beyond the SM. In particular,
the physics of large extra dimensions can be studied in $\gamma \gamma$
collisions at the NLC with the exciting possibility of discovering this
source of new physics, or (in the absence of any evidence for its existence)
to put stringent bounds on the parameters of this model. We study the
effects of virtual graviton exchange on the $t \bar t$ production cross-section
in $\gamma \gamma$ collisions with the above physics goal in mind. 

The simplest way to approach the photon-photon scattering process in these
experiments is to think of it as analogous to the parton model so
that the basic scattering is described by a $\gamma \gamma$ scattering
subprocess, with each $\gamma$ resulting from the electron-laser back
scattering. The energy of the back-scattered photon, $E_\gamma$, follows
a distribution characteristic of the Compton scattering process and
can be written in terms of the dimensionless ratio $x=E_\gamma/E_e$.
It turns out that the maximum value of $x$ is about 0.82 so that provides
the upper limit on the energy accessible in the $\gamma \gamma$ sub-process.
The subprocess cross-section is convoluted with
the luminosity functions, $f^i_\gamma (x)$, which provide information 
on the photon flux produced in Compton scattering of the electron
and laser beams \cite{lumino}. 

For the case of $t \bar t$ production that we consider in this paper,
the above discussion implies that we need only calculate the cross-section 
for the $\gamma \gamma \rightarrow t \bar t$ process and then convolute
this with the luminosity functions to get the full 
cross-section. The cross-section for the $\gamma \gamma \rightarrow t \bar t$ 
process has the usual $t$- and $u$-channel SM contributions, but in
addition, we also have the $s$-channel exchange of virtual spin-2
Kaluza-Klein particles. In the following, we refer to this contribution
as the non-SM (NSM) contribution. We calculate the cross-section for 
the polarised case and then obtain the unpolarised cross-section by 
summing over the polarisations of the initial photons. The polarisation 
of each of the photon is a function of the polarisations of the initial 
electron and laser beams and it is only the latter that can be fixed in 
the experiment. When we present our results for the polarised case, we will 
do so for a fixed choice of these initial polarisations. 

For the polarised case, the cross-section takes on the following factorised form
\begin{eqnarray}
d \sigma (\lambda_{e1},\lambda_{e2},\lambda_{\ell 1},\lambda_{\ell 2}) 
&=& \int d x_1 ~ \int d x_2 ~ f^1_\gamma (x_1,\xi) ~ 
                              f^2_\gamma (x_2,\bar\xi) 
\nonumber\\
	      && \times \left(
d\hat\sigma_{00}+\xi_2 ~\bar\xi_2 ~d\hat\sigma_{22}+
\xi_2 ~d\hat\sigma_{02}+ \bar\xi_2 ~d\hat\sigma_{20}
                 \right) ~d\Gamma ,
\end{eqnarray}
where $d\Gamma$ includes the phase space and flux factor.  $\xi$ and
$\bar \xi$ are the Stokes parameters of the first and second photons
respectively. For example, $f^1_\gamma(x_1,\xi)$ is the 
distribution of photons for a given polarisation $\lambda_{\ell 1}$ of 
the laser beam and a polarisation $\lambda_{e1}$ of the electron. 
$d \hat\sigma_{ij}$ is related to the polarised $\gamma ~\gamma 
\rightarrow t~ \bar t$ matrix elements as follows

\begin{eqnarray}
d\hat\sigma_{00} &=& \frac{1}{4} \sum_{\lambda_1 \lambda_2}
                 \vert M(\lambda_1,\lambda_2)\vert^2  ,
                 \nonumber\\
d\hat\sigma_{22} &=& \frac{1}{4} \sum_{\lambda_1 \lambda_2}
                 \lambda_1 \lambda_2 
                 \vert M(\lambda_1,\lambda_2)\vert^2  ,
                 \nonumber\\
d\hat\sigma_{20} &=& \frac{1}{4} \sum_{\lambda_1 \lambda_2}
                 \lambda_1 
                 \vert M(\lambda_1,\lambda_2)\vert^2  ,
                 \nonumber\\
d\hat\sigma_{02} &=& \frac{1}{4} \sum_{\lambda_1 \lambda_2}
                 \lambda_2 
                 \vert M(\lambda_1,\lambda_2)\vert^2 ,
                 \nonumber 
\end{eqnarray}
where $\lambda_{1,2}$ denote the circular polarisation of the photons 1,2
respectively.  

Summing over the polarisations of the electron and laser beam the only term
that survives in the above expression is $d \sigma_{00}$.  The unpolarised 
differential cross section can be written as
\begin{equation} 
\frac{d^3 \sigma} {dp_T^2 d~y} = \int d x_1 ~ \int d x_2 
~ f_\gamma (x_1) ~ f_\gamma (x_2) ~ 
\hat s \frac{d \hat \sigma} {d \hat t} ~
\delta (\hat s+\hat t+\hat u-2 m_t^2) ,
\end{equation} 
where the hatted variables corresponds to the $\gamma~ \gamma~ \rightarrow
t~ \bar t$ subprocess and $m_t$ is the mass of the top quark.  

\begin{eqnarray}
\frac{d \hat \sigma} {d \hat t} &=&
\left ({\frac{d \hat \sigma} {d \hat t}}\right)_{SM} +
\left ({\frac{d \hat \sigma} {d \hat t}}\right) _{NSM} ,
\end{eqnarray}

\begin{eqnarray}
\left({\frac{d \hat \sigma} {d \hat t}}\right)_{NSM} 
({\lambda_1,\lambda_2}) &=& -\frac{3}{2\hat s^2} [1-\lambda_1 \lambda_2] 
\left(
\frac {\pi \lambda^2} {M_s^8} -
\frac {4 \pi \alpha e_q^2 \lambda} {M_s^4} 
\frac{1}{(m_t^2-\hat t) (m_t^2-\hat u)} 
\right) 
\nonumber \\
&& \times \left\{ 2 m_t^8-8m_t^6 \hat t-2 m_t^2 \hat t(\hat s+2 \hat t)^2+
m_t^4 (\hat s^4+4 \hat s \hat t+12 \hat t^2) \right.
\nonumber \\
&&\left. + \hat t (\hat s^3+3 \hat s^2 \hat t+4 \hat s \hat t^2+2 
\hat t^3)\right \} .
\end{eqnarray}

Before we discuss the analysis that we have done, we would like to 
point out that the non-SM contribution involves two new parameters: 
the effective string scale $M_S$ and $\lambda$ which is the effective 
coupling at $M_S$. $\lambda$ is expected to be of ${\cal O}(1)$, but 
its sign is not known $a\ priori$. In our work we will explore the 
sensitivity of our results to the choice of the sign of $\lambda$. 

\vspace*{-2in}
\begin{figure}[tbhp]
   \begin{minipage}{60mm}
  \begin{flushright}
    \leavevmode
    \epsfig{bbllx=1,bblly=256,bburx=299,bbury=606,
        file=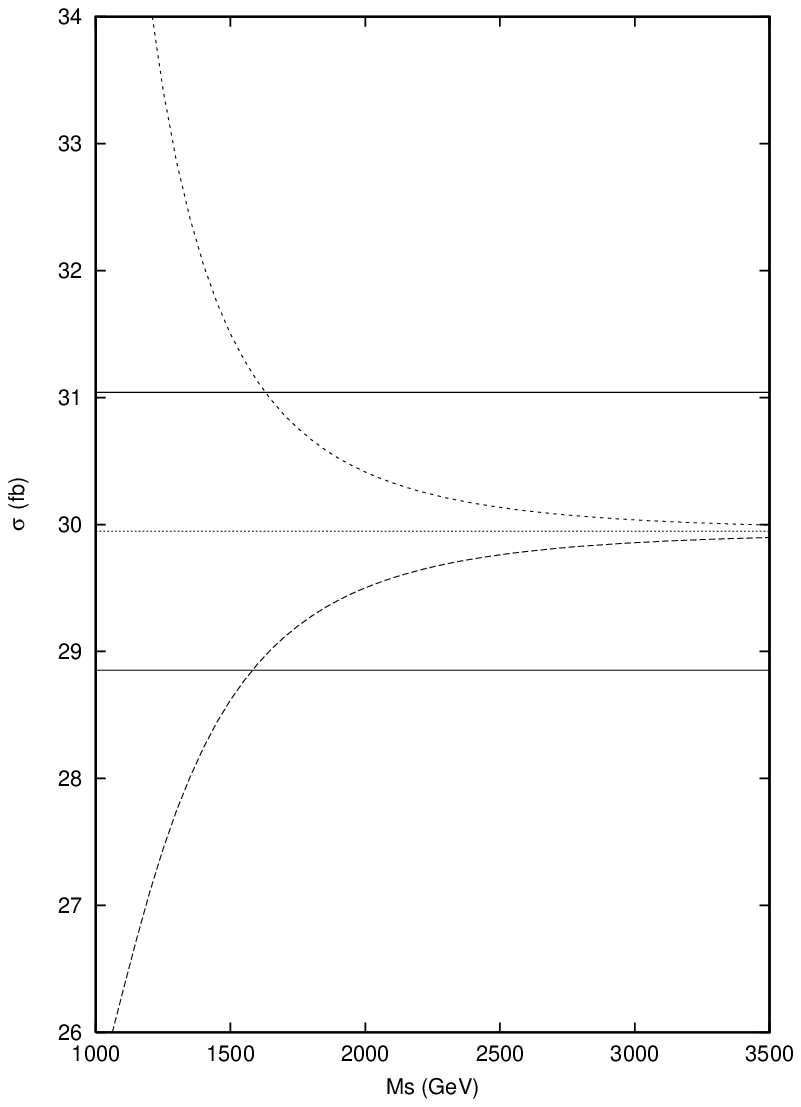,height=78mm,angle=0}
\label{fig-pdk1}
   \end{flushright}
   \end{minipage}
\hfill   
\begin{minipage}{60mm}
  \begin{flushright}
    \leavevmode
      \epsfig{bbllx=51,bblly=256,bburx=349,bbury=606,
          file=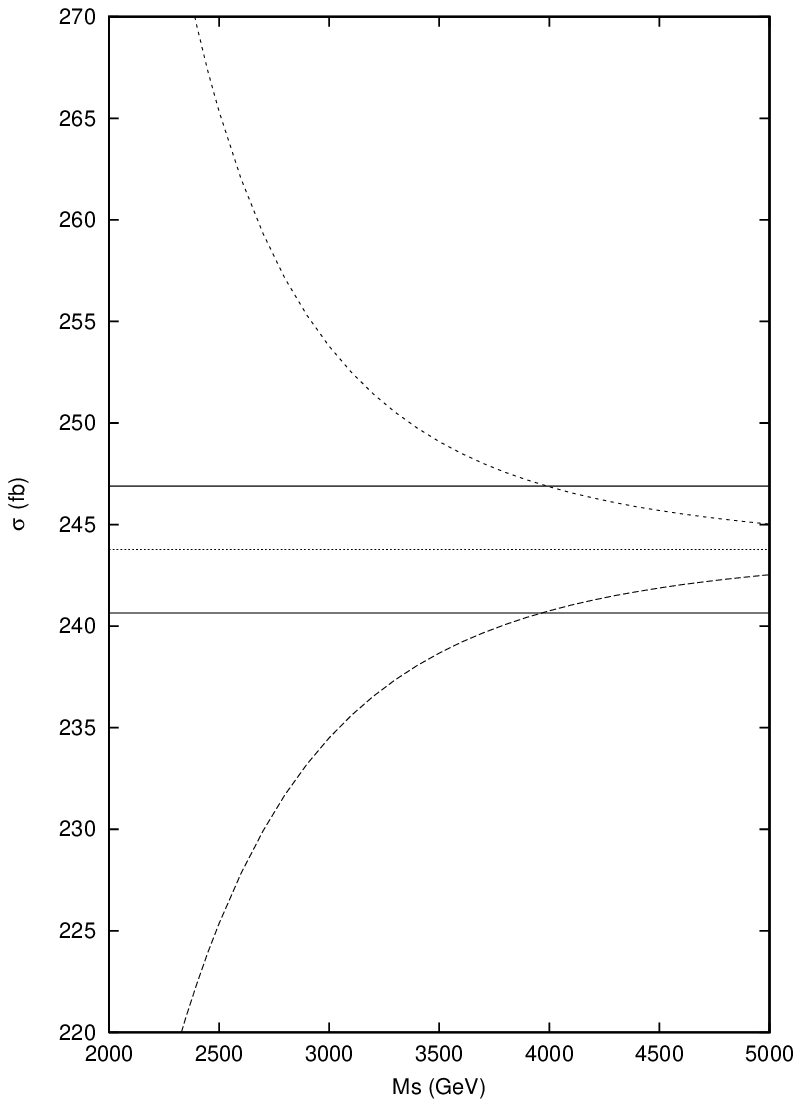,height=78mm,angle=0}
\label{fig-pdk2}
     \end{flushright}
   \end{minipage}
\hfill   
   \begin{minipage}{60mm}
  \begin{flushright}
    \leavevmode
    \epsfig{bbllx=91,bblly=256,bburx=389,bbury=606,
        file=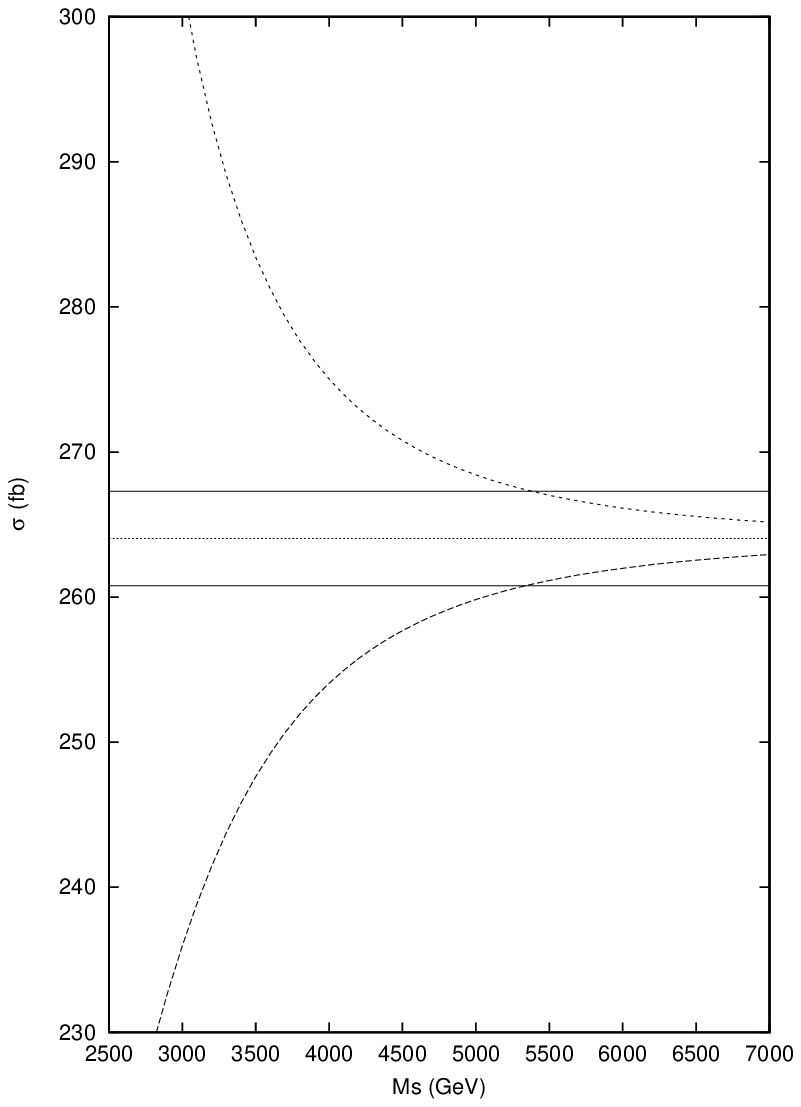,height=78mm,angle=0}
\label{fig-pdk3}
   \end{flushright}
   \end{minipage}
\end{figure}
\vskip150pt 
\begin{flushleft}
\noindent 
{\footnotesize \it Fig.1: $t \bar t$ production cross-section (in fb) as a function
of $M_S$ for $\sqrt{s}=$ 500 GeV (a), 1000 GeV (b) and 1500 GeV (c). The
upper (lower) curves are for $\lambda=-1(+1)$. The
lines show the SM value and the $2\sigma$ upper and lower limits.}
\end{flushleft}
 
We begin with the results for the unpolarised case. In Fig.~1, we
have plotted the unpolarised (integrated) cross-section as a function of $M_S$
for three different values of the initial $e^+ e^-$ C.M. energy
i.e. $\sqrt{s}= 500,\ 1000,\ 1500$~GeV (shown in Fig.~1 (a), (b)
and (c) respectively). Also shown in the figure are the SM value of
the cross-section and the $2\sigma$ upper and lower limits. For
obtaining the latter, we have assumed purely statistical errors
and assuming an integrated luminosity of 100 fb${}^{-1}$.
We find, first of all, that the curves are symmetrical with respect
to $\lambda$. The $2\sigma$ limits that we obtain for 
$\sqrt{s}= 500,\ 1000,\ 1500$~GeV are 1600, 4000 and 5400 GeV, respectively.
At the higher C.M. energies of $\sqrt{s}=1000,\ 1500$ GeV that we have
considered, the $t \bar t$ production cross-section is large enough
and with the integrated luminosity that we have assumed a large number
of $t \bar t$ events results. The integrated cross-section itself
becomes a good discriminator of the new physics effects.

In Fig.~2 we have plotted the rapidity distribution for $\sqrt{s}=1000$ 
GeV, in order to consider whether the use of differential quantities
will further enhance the sensitivity of the process under consideration
to the effects of the new physics. We find that the difference between
the SM and the SM$+$NSM distributions to be quite significant, especially
when we concentrate in the central regions of rapidity. Though we refrain
from making a quantitative estimate of the bound that would result (for
such an estimate would be premature without knowing further experimental 
details), it is clear from Fig.~2 that the rapidity distribution can
be used to improve the bound that would result from looking only at
the integrated cross-section.

\begin{figure}[htb]
\begin{center}
\epsfxsize=2.4in\leavevmode\epsfbox{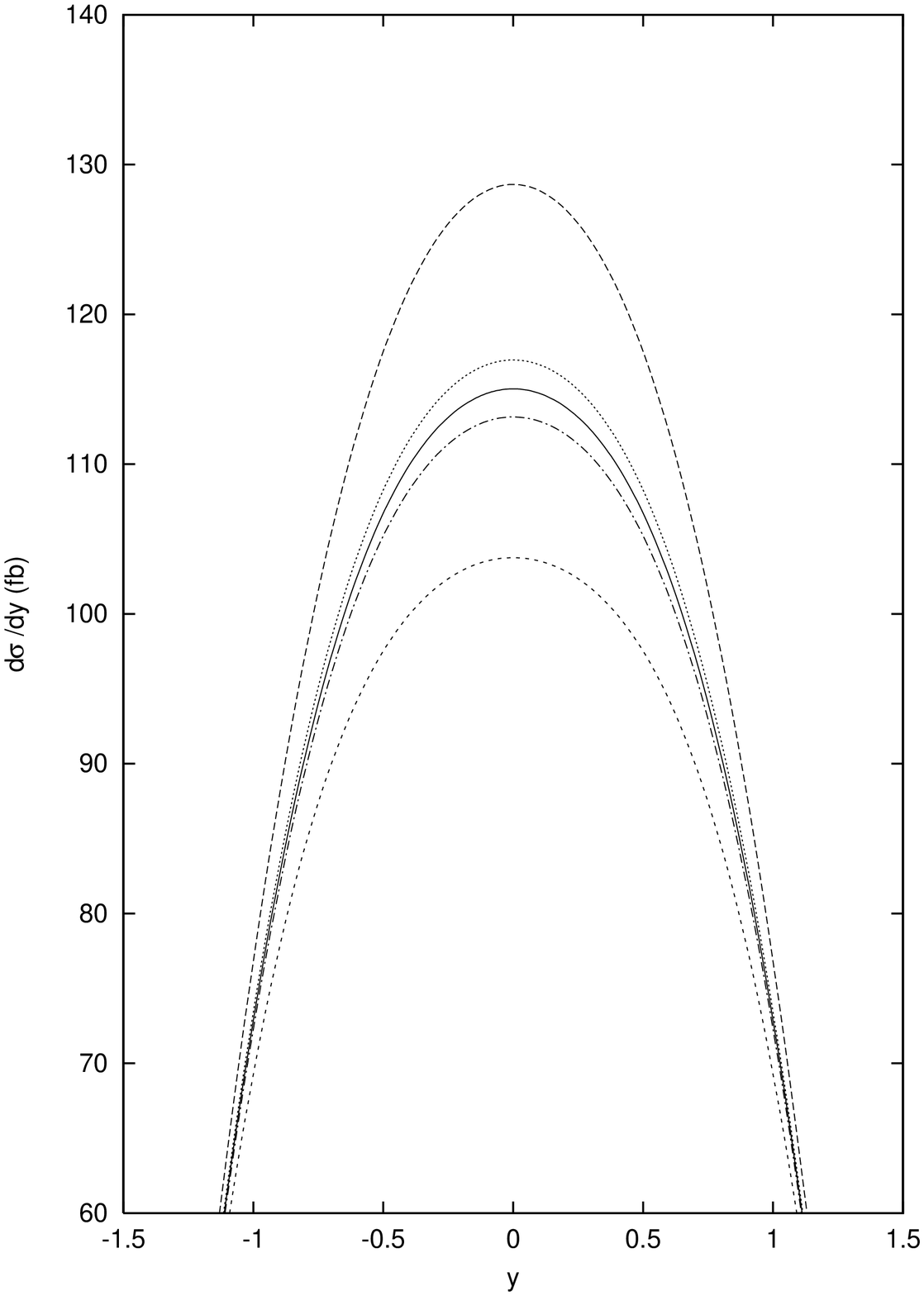}
\end{center}
\end{figure}
\begin{flushleft}
{\footnotesize\it Fig.2: The rapidity distribution for $\sqrt{s}=1000$ GeV. 
The solid curve is the SM distribution. The two curves below
the SM curve are for $M_S=1.2,\ 1.8$ TeV respectively and for $\lambda=1$,
while the curves above the SM curve for $\lambda=-1$. } 
\end{flushleft}

For the polarised case, we study the effects of the NSM contribution
for different choices of the initial electron and laser beams. For a
given choice of the $e^-$ and laser polarisation, the photon polarisation 
is fixed once the $x$ value is known. The latter polarisation is therefore
dependent crucially on the luminosity functions and it is only on the
polarisation of the electron and the laser beams that we have a direct
handle. The efficacy of polarisation as a discriminator of the new physics
is, however, apparent more at the level of the $\gamma \gamma$ sub-process.
As we scan over the different choices of initial beam polarisations, we
find that for certain choices there is hardly any sensitivity to the
new physics i.e. the SM and the SM$+$NSM cross-sections are more or
less the same. However, large differences are realised for certain other
choices $viz.$ for the cases $(+,-,-,-)$ and $(+,-,-,+)$, where these
represent the polarisations $(\lambda_{e1}, \lambda_{e2}, \lambda_{\ell 1},
\lambda_{\ell 2})$. These polarization combinations are very sensitive to 
the NSM contribution while other combinations are not so. This has to do 
with the fact that the NSM contribution (see eqn.6) is significant when 
$\lambda_1$ and $\lambda_2$ are with opposite signs. For hard photons, the 
sign of $\lambda$ is the same as that of the helicity of the initial 
electron. Therefore the combination that has $\lambda_1 \lambda_2$ $-$ve is 
the one with electron beams with opposite helicities $(\lambda_{e1} = -
\lambda_{e2})$. The luminosity function is such that, for the combination 
with $\lambda_e \lambda_l = -1$, it peaks in the high energy region. 
Therefore, the sensitivity of the combination with both the beams having 
$\lambda_{e1}\lambda_{l1}=-1=\lambda_{e2}\lambda_{l2}$ is the best while 
the combination with $\lambda_{e1}\lambda_{l1}=1=\lambda_{e2}\lambda_{l2}$ 
is least sensitive. The other combination has a sensitivity which is
intermediate. 

In Table 1, we show the $M_S$ limits for these polarisation
choices for the three different C.M. energies that we have considered. 
As can be clearly seen, the use of polarisation enhances the bounds on
$M_S$ quite significantly by several 100 GeV in each case.

\begin{center}
\begin{table}
\begin{center}
\begin{tabular}{|l|cl|cl|}

\hline
($\lambda_{e1}$ $\lambda_{e2}$ $\lambda_{\ell 1}$ $\lambda_{\ell 2}$)
  && $\sqrt s$ (GeV) && $M_s$ (GeV) \\
\hline
\hline

                          &&   500  &&   1950\\
(+$~~~$ -$~~~$ -$~~~$ - $~$)  &&  1000  &&  4600\\
                          &&  1500  &&  6000\\

\hline

                          &&   500  &&  2500\\
(+$~~~$ -$~~~$ -$~~~$ +)  &&  1000  &&  4800\\
                          &&  1500  &&  6400\\

\hline

\end{tabular}
\end{center}
\caption{$2\sigma$ limits on Ms (GeV) with polarized initial beams
$\lambda_{e1}$ $\lambda_{e2}$ $\lambda_{\ell 1}$ $\lambda_{\ell 2}$ are 
the helicities of initial electron and laser beams and 
$\sqrt{s}$ is the $e^-e^-$ centre of mass energy. An $e^-e^-$ Luminosity 
of $100 fb^{-1}$ is assumed.}
\end{table}
\end{center}

In summary, the effects of virtual spin-2 particle exchange in scenarios
with large extra Kaluza-Klein dimensions can be significantly tested in
$t \bar t$ production in $\gamma \gamma$ collisions at the NLC. The
unpolarised integrated rate in itself turns out to be a sensitive
discriminator for this scenario and the bounds on the effective
low-energy quantum gravity scale that result from the analysis of
the integrated rate is between approximately 1600~GeV and 5400~GeV
depending upon the initial $e^+ e^-$ C.M. energy. These bounds can
be strengthened by studying rapidity distributions or by tuning the
polarisations of the initial electron and laser beams.

\vskip25pt
{\sf Acknowledgments:} One of us (P.M.) would like to thank S. Umasankar
for hospitality at IIT, Mumbai where part of this work was done.

\end{document}